\documentclass[nofootinbib,aps,prd,groupedaddress,preprintnumbers,%
  showpacs,showkeys,floatfix,amssymb,amsfonts]{revtex4-1}  
\usepackage{diagbox}
\usepackage[usenames]{color}

\usepackage{ulem}
\usepackage{graphicx}
\usepackage{bm}
\usepackage{amsmath}
\usepackage{amssymb}
\usepackage{bbold}
\usepackage{amsfonts}
\usepackage{float}
\usepackage{dsfont}  
\usepackage{slashed}  
\usepackage{booktabs}
\usepackage{multirow}
\usepackage{subfigure}
\usepackage{slashed}
\usepackage[sort&compress]{natbib}

\usepackage{xr-hyper}

\usepackage[svgnames,x11names,table]{xcolor}
\usepackage[colorlinks=true,linkcolor=MediumBlue,citecolor=MediumBlue,urlcolor=MediumBlue]{hyperref}
\usepackage{hyperref}

\newcommand{\be}{\begin{equation}}
\newcommand{\ee}{\end{equation}}
\newcommand{\bea}{\begin{eqnarray}}
\newcommand{\eea}{\end{eqnarray}}

\def\smn{{\sigma_{\mu\nu}}}

\begin{document}

\title{Improved renormalization scheme for nonlocal operators}

\author{Martha Constantinou$^{1}$, Haralambos Panagopoulos$^2$ \footnote{Electronic address:\\ marthac@temple.edu,\\ haris@ucy.ac.cy}
\phantom{.}\\\phantom{.}}

\vspace*{0.75cm}
\affiliation{
$^1$ {\small\textit{Department of Physics,  Temple University,  Philadelphia,  PA 19122 - 1801,  USA}}\\\phantom{.}\\
$^2$ {\small\textit{Department of  Physics,  University  of Cyprus,  POB  20537,  1678  Nicosia,  Cyprus}}}

\begin{abstract}
\vspace*{1cm}

In this paper we present an improved RI-type prescription appropriate for the non-perturbative renormalization of gauge invariant nonlocal operators. In this prescription, the non-perturbative vertex function is improved by subtracting unwanted finite lattice spacing ($a$) effects, calculated in lattice perturbation theory. The method is versatile and can be applied to a wide range of fermion and gluon actions, as well as types of nonlocal operators. The presence of operator mixing can also be accommodated.

Compared to the standard RI' prescription, this variant can be recast as a supplementary finite renormalization, whose coefficients bring about corrections of higher order in $a$; consequently, it coincides with standard RI' as $a\to 0$, however it can afford us a smoother and more controlled extrapolation to the continuum limit.

In this proof-of-concept calculation we focus on nonlocal fermion bilinear operators containing a straight Wilson line. In the numerical implementation we use Wilson/clover fermions and Iwasaki improved gluons. The finite-$a$ terms were calculated to one-loop level in lattice perturbation theory, and to all orders in $a$, using the same action as the non-perturbative vertex functions. We find that the method leads to significant improvement in the perturbative region indicated by small and intermediate values of the length of the Wilson line. This results in a robust extraction of the renormalization functions in that region.

We have also applied the above method to operators with stout-smeared links. We show how to perform the perturbative correction for any number of smearing iterations, and evaluate its effect on the power divergent renormalization coefficients.

\end{abstract}

\maketitle

\section{Introduction}

The past decade has seen a great surge in the study of parton distribution functions (PDFs), by means of calculations formulated on a spacetime lattice. The viability of a Euclidean formulation for these intrinsically Minkowski-space functions was argued in ground-breaking papers by Ji~\cite{Ji:2013dva,Ji:2014gla} with the introduction of the Large Momentum Field Theory (LaMET) and the calculation of the so-called quasi-distributions. A number of approaches are currently being employed in addition to quasi-distributions in order to match experimentally measurable distribution functions to quantities amenable to numerical estimation on a lattice, such as the pseudo-distributions~\cite{Radyushkin:2017cyf,Orginos:2017kos}. Both of these cases rely on the evaluation, via lattice simulations, of matrix elements which involve gauge-invariant nonlocal composite operators.
There is a large variety of nonlocal operators which are of physical relevance, notably path-ordered exponentials of the gluon field along a straight-line or staple-shaped contour ("Wilson line"), with a quark-antiquark pair or two gluon field tensors attached at the endpoints. For an extended list of general references, we refer the reader to Refs~\cite{Monahan:2018euv,Cichy:2018mum,Ji:2020ect,Constantinou:2020pek,Cichy:2021lih}.

The study of nonlocal operators in Minkowski continuum spacetime has a long history, dating back decades before lattice investigations began, with seminal works by Mandelstam, Polyakov, Makeenko and Migdal \cite{Mandelstam:1968hz,Polyakov:1979gp,Makeenko:1979pb}, and later works, among them Refs.\cite{Witten:1989wf,Dotsenko:1979wb,Brandt:1981kf,Craigie:1980qs,Stefanis:1983ke,Dorn:1986dt,Chetyrkin:2003vi}.
The lattice formulation introduces several new complications, such as mixing and renormalization functions which exhibit an inverse-power divergence as the regulator $a$ (lattice spacing) approaches its limiting value, $a\to 0$. Power divergences can of course manifest themselves also in the renormalization of local operators, but only in the presence of mixing with lower-dimensional operators which share the same symmetry properties; nonlocal operators, on the other hand, by virtue of the natural length scale(s) appearing in their definition, exhibit power divergences despite the absence of mixing with lower-dimensional operators. Further, the effect of finite-$a$ corrections -- lattice artifacts -- becomes ever more significant as the spatial extent of the operator grows. 

A number of approaches are routinely applied to alleviate finite-$a$ artifacts. Many of them apply improved discretizations to terms in the action and to the definitions of composite operators; such improvements amount to adding contributions of higher order in $a$, which is a legitimate and systematic procedure in the spirit of the Symanzik-improvement program~(\cite{Symanzik:1983gh,Symanzik:1983dc}, for one of the first lattice applications see, e.g.,~\cite{Luscher:1996sc}). Exploitation of symmetries can also lead, in some cases, to elimination of ${\cal O}(a^1)$ (or even ${\cal O}(a^2)$\,) effects (e.g., \cite{Shindler:2007vp}, and \cite{Neuberger:1997fp,Luscher:1998pqa}). A recent approach developed for nonlocal operators, the hybrid scheme~\cite{Ji:2020brr}, implements a different renormalization scheme for small and large values of the length of the Wilson line, $z$. At short distances one may use an RI type or a ratio scheme. At large values of $z$ one can use the Wilson-line-mass-subtraction scheme~\cite{Chen:2016fxx,Ishikawa:2017faj}.

Another approach, which will be employed in the present work, has been quite successful in the renormalization of various local operators on the lattice; it relies on perturbation theory carried out to all orders in $a$ (see, e.g., Refs.~\cite{Capitani:2000xi,Constantinou:2009tr,Gockeler:2010yr,Constantinou:2013ada, Alexandrou:2015sea} for comparable implementations).
In this approach, the Green's functions which are to be evaluated non-perturbatively, in order to stipulate operator renormalization conditions, are also computed perturbatively, keeping the complete dependence on $a$ at the $i$-th perturbative order\footnote{Note that such contributions cannot be extracted analytically as functions of the external momentum $p$; rather, they must be computed numerically for every relevant value of $p$.
}, ${\cal O}(g^{2i})$. The perturbative result, generically denoted as $\Lambda^{(i)}(a)$, can be split into contributions which are ${\cal O}(a^0)$, ${\cal O}(a^0\ln(a))$ [possibly even ${\cal O}(a^{-1})$], and remaining contributions which vanish as $a\to 0$. This splitting is most straightforwardly performed to one loop: $\Lambda^{(1)}(a) = \Lambda^{(1)}_0(a) + \Lambda^{(1)}_{\rm rest}(a)$\,; the quantity $\Lambda^{(1)}_0(a)$ could be used to derive standard perturbative renormalization functions, should this be called for, while $\Lambda^{(1)}_{\rm rest}(a)$, which consists purely of lattice artifacts, can be subtracted from the non-perturbative value of the corresponding Green's function. The latter thus becomes a smoother function of $a$, and allows for a more stable extrapolation to the continuum limit. The legitimacy of this procedure can be best exposed by recasting the above subtraction in the form of a complementary renormalization containing only higher orders in $a$ [i.e., supplemental multiplicative renormalization functions of the form: $Z(a) = 1 + g^2 z(a),\ (z(0)=0)$ and mixing of the original operator $O$ with higher-dimensional operators $O_i$, thus leading to a renormalized operator $O^R = Z(a)\, O + \sum_i m_i(a)\, O_i,\ (m_i(0) = 0)$\,]. Ideally, the elimination of artifacts would circumvent the need to carry out an extrapolation to vanishing lattice spacing; this, unfortunately, is not practicable since, in the absence of such an extrapolation,
terms of ${\cal O}(g^{2i} a^n)$, ($i>1, n>0$), which are beyond computational feasibility, would still be significant  for typical values of $a$ used in simulations.

In this work, we apply this last approach to nonlocal operators; we focus on straight Wilson lines ending on a quark and an antiquark field.
It is worth noting that, in handling nonlocal operators, elimination of lattice artifacts is performed separately for different values of the operators' length scale(s), since each value corresponds to an independent operator. One of the advantages of this method is that it can be integrated in the hybrid scheme to improve the estimates for the small-$z$ region.

Another aspect of nonlocal operator improvement which is addressed in this work regards stout smearing~\cite{Morningstar:2003gk} and, in particular, the elimination of lattice artifacts from Green's functions of operators containing stout links. Implementation of stout smearing is straightforward and can be iterated an arbitrary number of times (``smearing steps") in the context of a numerical simulation. On the contrary, perturbative calculations in the presence of stout links are notoriously cumbersome, and can rarely be carried out beyond one step~\cite{Alexandrou:2016ekb}. We demonstrate that, for the operators under study, as well as for a rather wide class of local and nonlocal operators, one-loop perturbative expressions can be obtained as closed form integrands over the loop momentum, for an arbitrary number of smearing steps. This allows a perturbative treatment of artifacts, using the procedure outlined above.

The layout of this paper is as follows: Section II provides the general setup and definitions of the objects entering the calculation. In Section III we describe the renormalization scheme and the improvement procedure. Sections IV and V contain our perturbative results to all orders in $a$, and their effect of non-perturbative renormalization functions, respectively. Section VI discusses stout smearing and the results of its application to one loop, for an arbitrary number of smearing steps. Finally, Section VII summarizes our findings.

\section{Formulation}

\subsection{Lattice Actions}

We calculate the Green's functions of nonlocal operators using the clover (Sheikholeslami-Wohlert) fermion action~\cite{Sheikholeslami:1985ij}
\bea
S_F=
&-&
\frac{a^3}{2}\sum_{x,\,f,\,\mu}\bigg{[}\bar{
 \psi}_{f}(x) \left( r - \gamma_\mu\right) U_{x,\, x+a\,\mu}\psi_f(x+a\,\mu) 
+\bar{\psi}_f(x+a\,\mu)\left( r + \gamma_\mu\right)U_{x+a\,\mu,\,x}\psi_{f}(x)\bigg{]}\nonumber \\
&+&
a^4 \sum_{x,\,f} (\frac{4r}{a}+m^f_0)\bar{\psi}_{f}(x)\psi_f(x) 
-\frac{a^5}{4}\,\sum_{x,\,f,\,\mu,\,\nu} c_{\rm SW}\,\bar{\psi}_{f}(x)
\smn {\hat F}_{\mu\nu}(x) \psi_f(x)\,,
\label{clover}
\eea
in which the clover parameter, $c_{\rm SW}$, is treated as a free parameter. The Wilson parameter, $r$, is set to 1; $f$ is a flavor index, $\smn=[\gamma_\mu,\,\gamma_\nu]/2$ and ${\hat F}_{\mu\nu}/(i a^2 g)$ is a clover discretization of the gluon field tensor, defined as
\be
{\hat F}_{\mu\nu} \equiv \frac{1}{8}\,(Q_{\mu\nu} - Q_{\nu\mu})\,,
\ee
with $Q_{\mu\nu}$ being\footnote{For ease of notation, we will often omit $a$ in what follows.}
\begin{eqnarray}
Q_{\mu\nu} \hspace{-0.02cm}&=& \hspace{-0.02cm} U_{x,\, x+\mu}U_{x+\mu,\, x+\mu+\nu}U_{x+\mu+\nu,\, x+\nu}U_{x+\nu,\, x}
+ U_{ x,\, x+ \nu}U_{ x+ \nu,\, x+ \nu- \mu}U_{ x+ \nu- \mu,\, x- \mu}U_{ x- \mu,\, x} \nonumber \\
&+&\hspace{-0.02cm} U_{ x,\, x- \mu}U_{ x- \mu,\, x- \mu- \nu}U_{ x- \mu- \nu,\, x- \nu}U_{ x- \nu,\, x}
+ U_{ x,\, x- \nu}U_{ x- \nu,\, x- \nu+ \mu}U_{ x- \nu+ \mu,\, x+ \mu}U_{ x+ \mu,\, x}\,.
\end{eqnarray}
The Lagrangian masses for each flavor, $m^f_0$, are set to their critical value, which to one-loop level is zero.

\bigskip
We present results for the Symanzik-improved gluon action~\cite{Horsley:2004mx}
\be
S_G=\frac{2}{g_0^2} \Bigl[c_0 \sum_{\rm plaq.} {\rm Re\,Tr\,}\{1-U_{\rm plaq.}\}
\,+\, c_1 \sum_{\rm rect.} {\rm Re \, Tr\,}\{1- U_{\rm rect.}\} 
+ c_2 \sum_{\rm chair} {\rm Re\, Tr\,}\{1-U_{\rm chair}\} \Bigr]\,,
\label{Symanzik}
\ee
with $c_0=3.648$, $c_1=-0.331$, and $c_2$ defined by the normalization condition $c_0 + 8 c_1 + 16 c_2 = 1$. We will focus on the Iwasaki action, for which $c_2=0$. Our calculation is readily applicable for a family of Symanzik improved gluon actions, as defined in Ref.~\cite{Horsley:2004mx}.

\subsection{Definition of Operators}

We study fermion bilinear nonlocal operators of the form
\be
\mathcal{O}_\Gamma\equiv \overline\psi(x)\,\Gamma\,\mathcal{P}\, 
e^{i\,g\,\int_{0}^{z} A_\mu(x+\zeta\hat\mu) d\zeta}\, \psi(x+z\hat{\mu})\,,
\label{Oper}
\ee
where the path-ordered (${\cal P}$) Wilson line is inserted for gauge invariance. We consider only cases where the Wilson line is straight, and, without loss of generality, we choose it to be along the ${\hat z}$ direction. The length of the Wilson line is $z$, and our perturbative calculation implements values of $z$ which range between $z=0$ and $z=15$. This range is sufficient for the lattice sizes used currently in non-perturbative calculations of the renormalization functions. Note that, in the limit $z\to 0$, Eq.~(\ref{Oper}) reduces to the standard ultra-local fermion bilinear operators. However, the calculation of the Green's functions for $\mathcal{O}_\Gamma$ is for strictly $z\neq 0$ as the limit $z\to 0$ is nonanalytic.

We perform our calculation for all 16 independent combinations of Dirac matrices, $\Gamma$, and we distinguish between the cases in which a Lorentz index of a Dirac matrix is in the direction $z$ of the Wilson line or not. Therefore, the possible choices of $\Gamma$ are separated into 8 subgroups, defined as
\bea
\label{Oper2}
  {S} \equiv \mathcal{O}_{\mathbb{1}}\,,  \quad\,\,       
  {P} \equiv \mathcal{O}_{\gamma^5}\,,   \quad &&     
  {V}_z \equiv \mathcal{O}_{\gamma^z}\,,   \quad\,\,      
  {V}_j \equiv \mathcal{O}_{\gamma^j}\,,   \nonumber \\[0.5ex] 
  {A}_z \equiv \mathcal{O}_{\gamma^5\gamma^z}\,,  \quad\,\,  
  {A}_j \equiv \mathcal{O}_{\gamma^5\gamma^j} \,,  \quad&&
  {T}_{zj} \equiv \mathcal{O}_{\sigma^{zj}} \,,   \quad\,\, 
  {T}_{jk} \equiv \mathcal{O}_{\sigma^{jk}}\,, 
\eea
where $j, k \in \{t,\,x,\,y\}$. An alternative definition for the tensor operators is $\mathcal{O}_{\gamma^5\sigma^{\mu\nu}}$\,. This is redundant if one employs a
4-dimensional regularization, such as the lattice, since the latter operators are just a renaming of the $T$ operators, and they will, thus, renormalize identically.

Ref.~\cite{Constantinou:2017sej} revealed a pairwise finite mixing pattern in lattice regularization for some of the operators of Eq.~\eqref{Oper2}, based on one-loop perturbative calculations in lattice QCD. The mixing pattern is in the pairs $\{S,\, V_z\}$, $\{A_t,\, T_{xy}\}$, $\{A_x,\, T_{ty}\}$, $\{A_y,\, T_{tx}\}$, and this holds to all orders in perturbation theory. The improvement method that we propose here can be applied to the cases with or without  mixing.

\subsection{Perturbative Calculation}

The central focus of this work is the evaluation of the lattice-regularized bare amputated Green's functions 
\be
\label{eq:Green}
\Lambda^{\rm 1-loop}_{\cal O} = \langle \psi\,{\cal O}_{\Gamma}\,\bar \psi \rangle^{\rm 1-loop}\,,
\ee 
to one-loop level in perturbation theory and all orders in the lattice spacing, ${\cal O}(g^2\,a^\infty)$.
Exploiting translational invariance in this equation, the operator ${\cal O}_\Gamma$ is summed over all space-time positions, and thus the external quark and antiquark Fourier transformed fields have the same momentum ($p$).
$\Lambda^{\rm 1-loop}_{\cal O}$ explicitly depends on $p$ and on the length of the Wilson line $z$. Such a calculation requires more sophisticated methods of integration over the loop momentum, but is, overall, less complicated than the calculation to ${\cal O}(g^2\,a^0)$, which must also be carried out. Arriving at analytic expressions for the latter necessitates that the divergent terms be isolated; this is a rather delicate procedure involving special techniques~\cite{Constantinou:2017sej}. On the contrary, the Green's functions ${\cal O}(g^2\,a^\infty)$ are evaluated numerically for each value of the external momentum that is used in the numerical calculation of the renormalization functions; this momentum serves as the renormalization scale. Besides these 1-loop calculations, we evaluate non-perturbatively $\Lambda^{\rm 1-loop}_{\cal O}$, traced with its tree-level value.
Such a projection, described in more detail in Sec.~\ref{sec:renorm}, eliminates the Dirac structure, leading to a computationally less costly calculation. Other kinds of projectors can also be employed.

\begin{figure}[ht!]
\centerline{\includegraphics[scale=1]{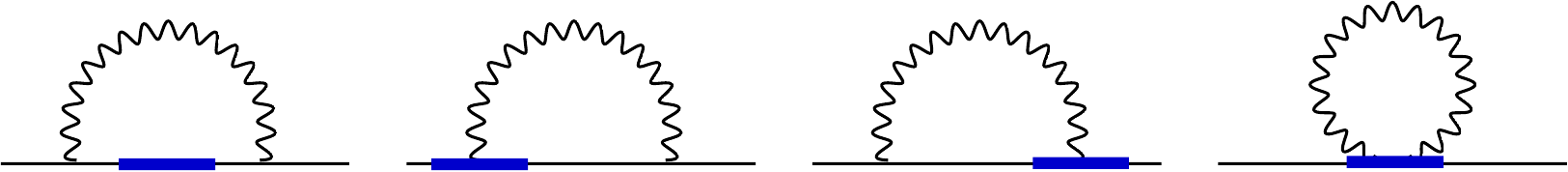}}
\centerline{\bf{d1\hspace*{3.5cm} d2\hspace*{4cm} d3\hspace*{3.5cm} d4}}
\vspace*{-0.3cm}
\begin{center}
\begin{minipage}{15cm}
\hspace*{3cm}
\caption{\small{Feynman diagrams contributing to the one-loop calculation of the Green's functions of operator $\mathcal{O}_\Gamma$.
The straight (wavy) lines represent fermions (gluons). The operator insertion is denoted by a filled rectangle.}}
\label{fig1}
\end{minipage}
\end{center}
\end{figure}

The one-loop Feynman diagrams which enter our calculation are shown in Fig.~\ref{fig1}. The filled rectangle shows the insertion of any one of the nonlocal operators $\mathcal{O}_\Gamma$. Diagram d1 contains the 0-gluon vertex of the operator, whereas diagrams d2-d3 (d4) contain the corresponding 1-gluon (2-gluon) vertex. The ``tadpole'' diagram d4 is related to the self-energy of the operator and leads to the well-known linear divergence with respect to the dimensionful ultraviolet cutoff $a$~\cite{Dotsenko:1979wR,Brandt:1981kf}. As noted in Ref.~\cite{Constantinou:2017sej}, such a divergence is independent of the choice of the operator, and, to ${\cal O}(g^2\,a^0)$, has the form $c\,|z|/a$. The coefficient $c$ depends, however, on the gluon action. 

\section{Renormalization}
\label{sec:renorm}

\subsection{Non-perturbative calculation}

For the renormalization condition we employ a regularization independent RI$'$ scheme~\cite{Martinelli:1994ty}, which requires the calculation of the non-perturbative vertex functions, $G_{\cal O}(p,z)$. The latter correspond to a nonlocal operator within quark states. A convenient choice is $\langle\bar{\psi}_u\, {\cal O}_\Gamma \, \psi_d\rangle$, where $\psi_u$ ($\psi_d$) is the up-quark (down-quark) field. The vertex functions with momentum $p$ are amputated using the up-quark and down-quark propagator in momentum space, that is,
\begin{equation}
   {\cal V}_{\cal O}(p,z) = (S_u(p))^{-1}\, G_{\cal O}(p,z)\, (S_d(p))^{-1} \, .
\label{vertexfunction}
\end{equation}
${\cal V}_{\cal O}(p,z)$ is the non-perturbative equivalent of $\Lambda^{\rm 1-loop}_{\cal O}$, cf. Eq.~(\ref{eq:Green}). 
This amputated vertex function is matched to its tree-level value, ${\Lambda}_{\cal O}^{\rm tree}(p,z)$, using the RI$'$ prescription, where the vertex momentum is set equal to the renormalization scale. We define the renormalization functions for the quark field and quark operator as $\psi^R = Z_q\, \psi^B$ and ${\cal O}_\Gamma^R = Z_{\cal O} \,{\cal O}_\Gamma^B$, respectively. The appropriate condition for the renormalization function, $Z_{\cal O}$, is 
\be
\label{renormZO}
 Z_{\cal O}(\mu_0,z) = \frac{Z_q(\mu_0)}{\frac{1}{12} {\rm Tr} \left[{\cal V}_{\cal O}(p,z) \left(\Lambda_{\cal O}^{\rm tree}(p,z)\right)^{-1}\right]} \Bigr|_{p^2{=}\mu_0^2}\, ,
\ee
where the quark field renormalization, $Z_q$, is given by matching the propagator to its tree level value $\Lambda_q^{\rm tree}(p)$
\be
\label{renormZq}
Z_q(\mu_0) = \frac{1}{12} {\rm Tr} \left[S^{-1}(p)\, \Lambda_q^{\rm tree}(p)\right] \Bigr|_{p^2=\mu_0^2}\,.
\ee
Eq.~(\ref{renormZO}) is a generalization of the condition used for local operators at a scale $\mu_0$; here it is applied at each value of $z$ separately.

\subsection{Improvement scheme}

The improvement procedure is applied on Eq.~\eqref{renormZO} using the perturbative expressions for the ${\cal O}(g^2\,a^\infty)$ results obtained through the diagrams of Fig.~\ref{fig1}. A similar improvement is applied on $Z_q$ following the work of Ref.~\cite{Alexandrou:2015sea}. More precisely, we redefine the renormalization functions as follows: 
\be
\label{renormZO_impr}
 Z^{\rm impr}_{\cal O}(\mu_0,z) = 
 \frac{Z^{\rm impr}_q(\mu_0)}{\frac{1}{12} \left(
 {\rm Tr} \left[{\cal V}_{\cal O}(p,z) \left(\Lambda_{\cal O}^{\rm tree}(p,z)\right)^{-1}\right] - 
 {\rm Tr} \left[\Delta\Lambda^{\rm 1-loop}_{\cal O}(p,z) \left(\Lambda^{\rm tree}_{\cal O}(p,z)\right)^{-1}\right] 
 \right)} \,\Biggr|_{p^2{=}\mu_0^2} = 1\, ,
\ee
\be
\label{renormZq_impr}
Z^{\rm impr}_q(\mu_0) = \frac{1}{12} \Bigg(
{\rm Tr} \left[S^{-1}(p)\, \Lambda_q^{\rm tree}(p)\right] -
{\rm Tr} \left[(\Delta\Lambda_q^{\rm 1-loop}(p))^{-1}\, \Lambda_q^{\rm tree}(p)\right] 
\Bigg) \, \Biggr|_{p^2=\mu_0^2}\,,
\ee
where $\Delta\Lambda$ is the difference between the Green's functions to ${\cal O}(g^2\,a^\infty)$ and the Green's functions to ${\cal O}(g^2\,a^0)$, defined above as $\Lambda^{(1)}_{\rm rest}(a)$. With this improvement scheme, one subtracts the unwanted finite-$a$ contamination from the non-perturbative trace of the vertex functions using the results from 1-loop perturbation theory. There are alternative subtraction schemes, such as to subtract the finite-$a$ contamination on $Z_{\cal O}$. The latter differs from the one in Eq.~\eqref{renormZO_impr} to ${\cal O}(g^4 a^n)$ ($n\ge1$). As described in the Introduction, the variations of improvement schemes can also be understood as a finite renormalization whose effects disappear in the limit $a \to 0$. 

Once the improvement procedure is applied on the renormalization conditions, the analysis proceeds as outlined in Ref.~\cite{Alexandrou:2019lfo}, that is
\begin{itemize}
\item[\textbf{1.}] Calculation of the vertex functions using isotropic (``democratic'') momenta in the spatial directions, $ap=\frac{2\pi}{L} (\frac{n_t}{2} + \frac{1}{4},n_x,n_x,n_x)$. We employ antiperiodic boundary conditions in the time direction. In addition, the momenta should have a ratio
\be
P4 \equiv \frac{\sum_i(ap_i)^4}{(\sum_i(ap_i)^2)^2}
\ee
as close to 0.25 as possible (ideally below 0.3 for nonlocal operators). This is based on empirical arguments, since the Lorentz non-invariant function $P4$ appears in ${\cal O}(g^2 a^2)$ contributions in lattice perturbation theory.
\item[\textbf{2.}] Implementation of an improvement scheme that eliminates finite-$a$ contamination, such as Eq.~\eqref{renormZO_impr}.
\item[\textbf{3.}] Chiral extrapolation using the estimates of $Z_{\cal O}$ obtained from ensembles with different values for the quark masses.
\item[\textbf{4.}] Conversion of the chirally extrapolated $Z_{\cal O}$ to the modified $\overline{\rm MS}$ scheme (${\rm M}\overline{\rm MS}$) scheme~\cite{Alexandrou:2019lfo} and evolution to a common renormalization scale. The appropriate expressions for nonlocal operators can be found in Ref.~\cite{Constantinou:2017sej}.
\item[\textbf{5.}] Elimination of residual dependence on the initial renormalization scale using a linear fit in $(a\mu_0)^2$.
\end{itemize}

\section{Perturbative results}
\label{sec:results_pert}

It is instructive to study the finite-$a$ contributions extracted from our perturbative calculation, in order to assess the magnitude of the artifact contamination in the renormalization functions. For demonstration purposes we show the artifacts in the Landau gauge for $c_{\rm SW}=1$, and lattice size $L=24$. For the bare coupling constant ($g_0^2=6/\beta$) we choose $\beta=2.10$. These parameters match those of the calculation presented in Ref.~\cite{Alexandrou:2019lfo}. We note that the choice of the optimal tree-level value $c_{\rm SW}=1$ is consistent with 1-loop perturbation theory, and not with the value used in non-perturbative calculations. Using $c_{\rm SW}=1$ in perturbative calculations is the standard treatment of the clover parameter.

The one-loop lattice artifacts are embodied in the following expression:
\be
T_{\cal O}(z,p) \equiv \frac{1}{4} {\rm Tr} \left[\Delta\Lambda^{\rm 1-loop}_{\cal O}(p,z) \left(\Lambda^{\rm tree}_{\cal O}(p,z)\right)^{-1}\right] \,,
\ee
where 
\be
\Delta\Lambda^{\rm 1-loop}_{\cal O} = \Lambda^{\rm 1-loop,\, {\cal O}(g^2\,a^\infty)}_{\cal O} - \Lambda^{\rm 1-loop, \,{\cal O}(g^2\,a^0)}_{\cal O}\,.
\ee
The factor of $\frac{1}{4}$ is the normalization assuming that the trace is over the Dirac indices. We note that $T_{\cal O}(z,p)$ is a complex function due to the presence of the Wilson line. Numerically, we find that $T_{\cal O}$ is similar for all operators, and has mild $c_{\rm SW}$ dependence. Therefore, we focus on the vector operator, $\Gamma=\gamma^0$, for $c_{\rm SW}=1$. For simplicity, we drop the argument $p$ in $T_{\cal O}$.
We use several values of the momentum, $(ap)$, that correspond to $\{n_t,n_x\} = \{[0-10],[0-6]\}$, leading to a total of 77 values of the momentum in the range, $(ap)^2 \in [0, 10]$. The values of $P4$ cover the whole range, that is $P4 \in [0.25,1]$.

\begin{figure}[ht!]
    \centering
    \includegraphics[scale=0.47]{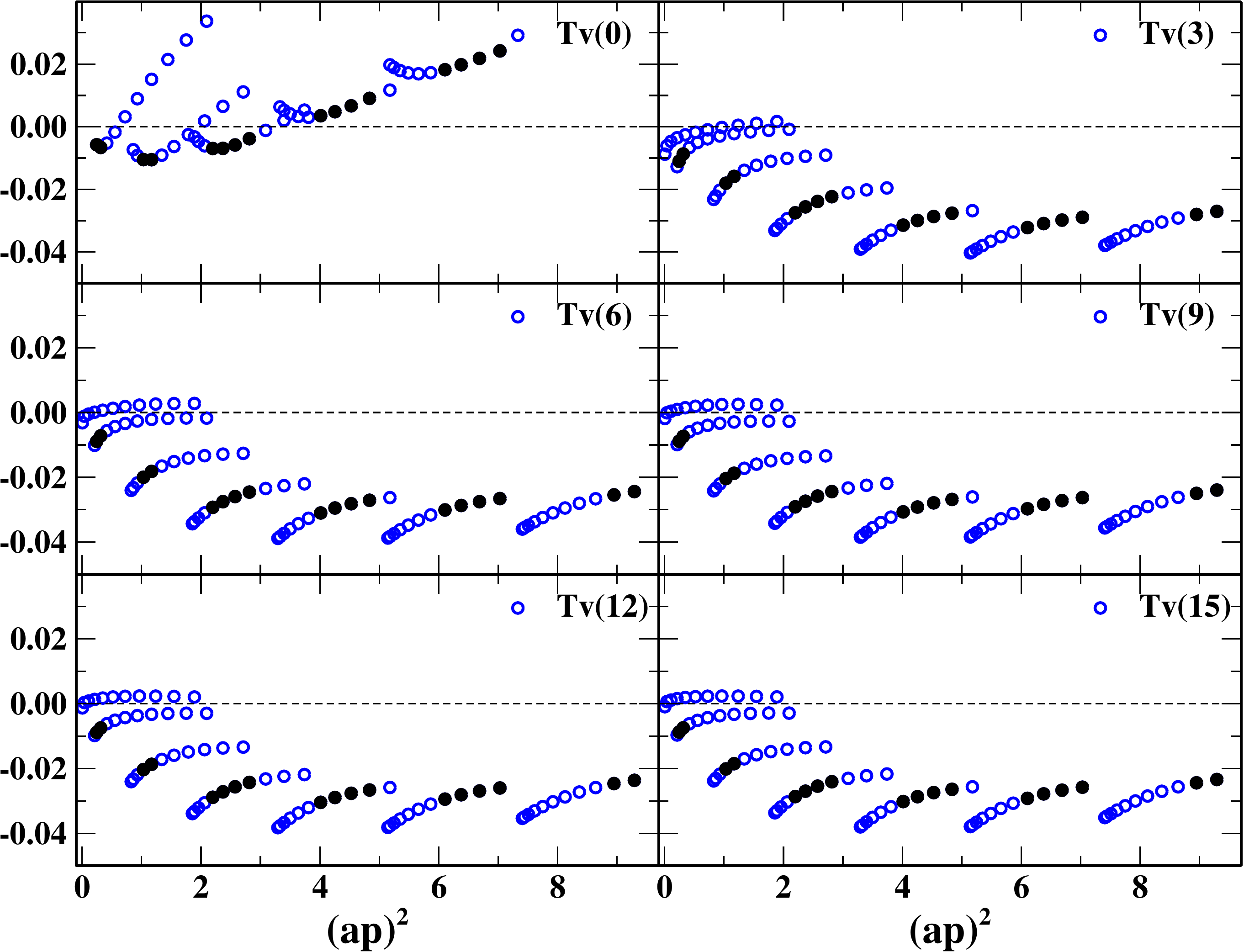}
    \caption{The real part of $T_V$ in the Landau gauge for $\beta=2.10$ and $c_{\rm SW}=1$ as a function of $(ap)^2$. We use $z=0,\,3,\,6,\,9,\,12,\,15$. The filled black symbols correspond to $(ap)$ values that have $P4\lesssim 0.26$.}
    \label{fig:Re_Artifacts_Tv}
\end{figure}

\begin{figure}[ht!]
    \centering
    \includegraphics[scale=0.47]{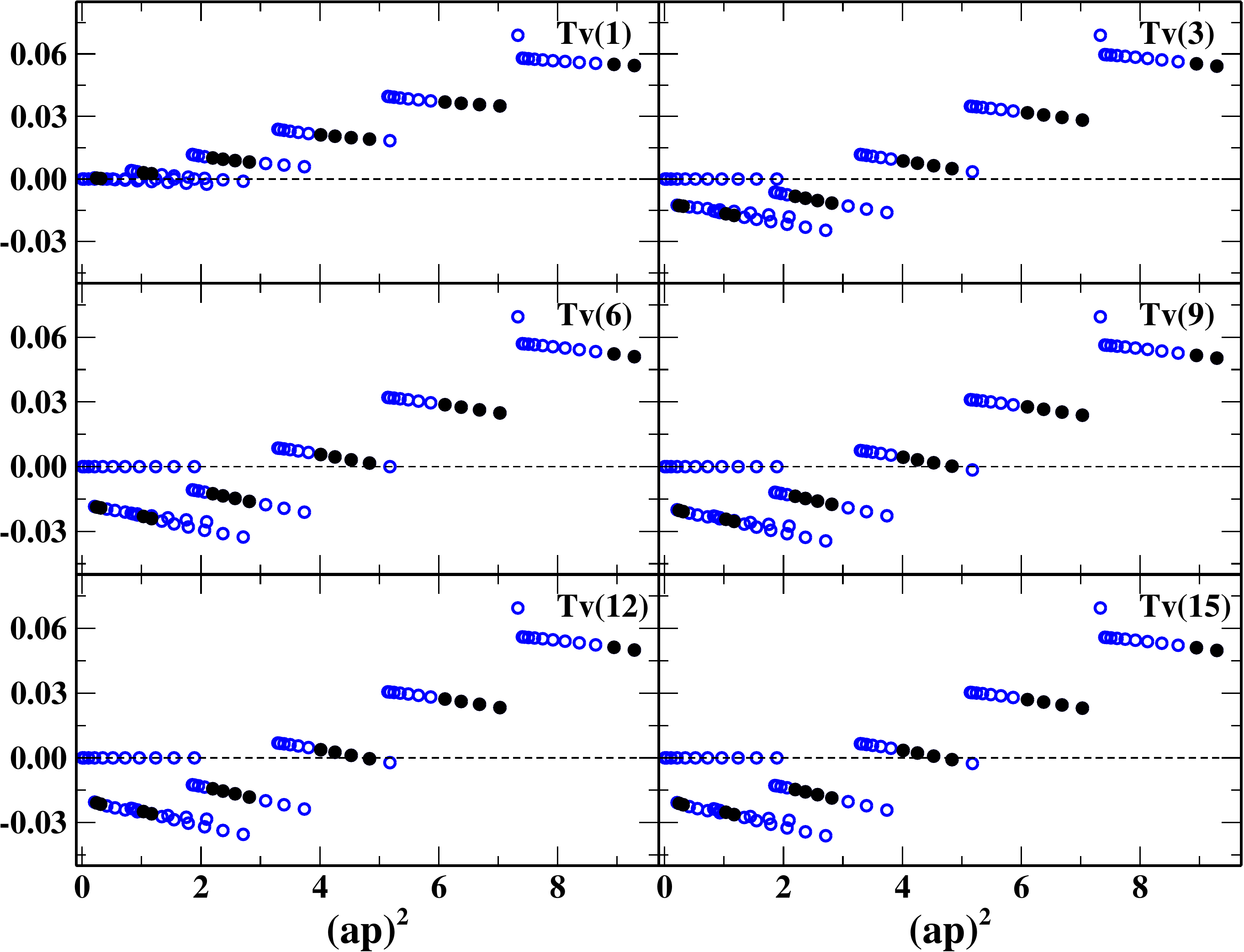}
    \caption{The imaginary part of $T_V$ in the Landau gauge for $\beta=2.10$ and $c_{\rm SW}=1$ as a function of $(ap)^2$. We use $z=1,\,3,\,6,\,9,\,12,\,15$. The filled black symbols correspond to $(ap)$ values that have $P4\lesssim 0.26$.}
    \label{fig:Im_Artifacts_Tv}
\end{figure}

Fig.~\ref{fig:Re_Artifacts_Tv} shows the real part of $T_V$ for selected values of the length of the Wilson line, namely, $z=3,\,6,\,9,\,12,\,15$. For comparison, we also include the values of $T_V(0)$ (top left panel), which are obtained from the calculation of Ref.~\cite{Alexandrou:2015sea}. Similarly, Fig.~\ref{fig:Im_Artifacts_Tv} corresponds to the imaginary part of $T_V$ for $z=1,\,3,\,6,\,9,\,12,\,15$. Given that $T_{\cal O}$ is purely real for $z=0$, we include $z=1$ instead. For both parts, we observe that the seven classes of momenta in the spatial direction ($n_x \in [0, 6]$) contributing to the same $z$ have a different behavior compared to each other that exhibits discontinuities. This holds for any value of $z$, including $z=0$. Such a ``wiggling'' effect has also been observed in the non-perturbative renormalization of local and nonlocal operators~\cite{Alexandrou:2015sea, Alexandrou:2019lfo}. On the contrary, the members of each class of $n_x$  ($n_t \in [0, 10]$) follow a smooth functional form, a feature also observed in non-perturbative calculations.

\begin{figure}[ht!]
    \centering
    \includegraphics[scale=0.47]{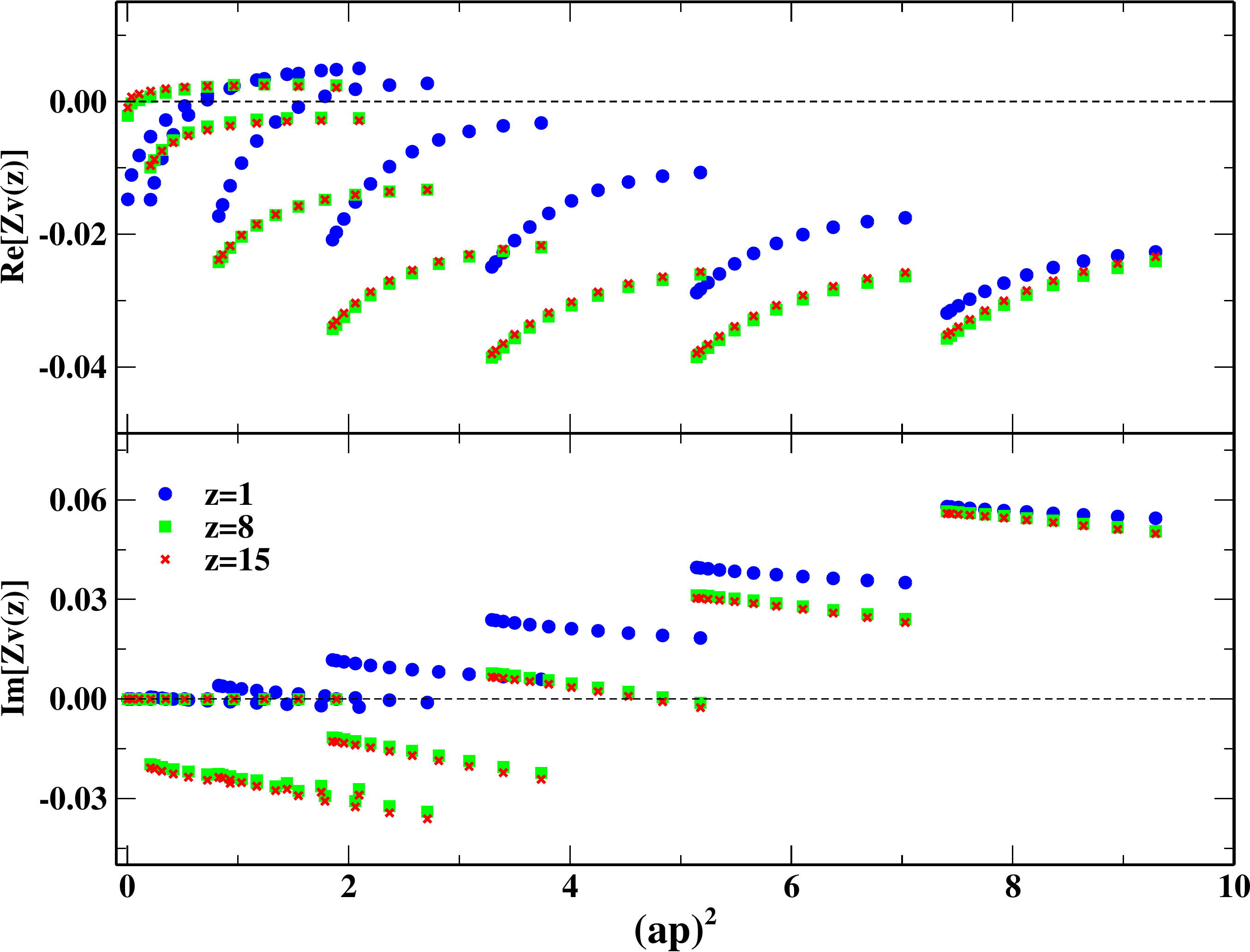}
    \caption{The real (top) and imaginary (bottom) parts of $T_V$ in the Landau gauge for $\beta=2.10$ and $c_{\rm SW}=1$ as a function of $(ap)^2$. Blue, green and red symbols correspond to $z=1,\,8,\,15$, respectively.}
    \label{fig:Artifacts_Tv_z1z8z15}
\end{figure}

There are a number of observations regarding the behavior of the finite-$a$ terms. We find that ${\rm Re}[T_V]$ is negative for $z>0$ for all momentum classes except the ones with spatial components in the class $(n_t,0,0,0)$. The artifacts for both ${\rm Re}[T_V](0)$ and ${\rm Im}[T_V]$ have negligible values for the class $(n_t,0,0,0)$. Another observation is that for $n_x>0$ the values of ${\rm Re}[T_V](0)$ are positive, while for $z>0$ the contributions are negative. Furthermore, ${\rm Re}[T_V](z>0)$ has a similar range of values, $[0,-0.04]$ for all values of $z$ shown. On the contrary, ${\rm Im}[T_V](z>0)$ has both positive and negative values that span the interval $[-0.04,0.06]$. The interplay between real and imaginary part of $T_V$ has nontrivial implications on Eq.~\eqref{renormZO_impr} due to its complex nature and will be studied numerically in Sec.~\ref{sec:results_nonpert}.
For convenience, in Fig.~\ref{fig:Artifacts_Tv_z1z8z15} we directly compare the artifacts of Figs.~\ref{fig:Re_Artifacts_Tv} - \ref{fig:Im_Artifacts_Tv} for $z=1,\,8,\,15$, in order to illustrate the $z$-dependence of the finite-$a$ contributions.

\section{Non-perturbative improved results}
\label{sec:results_nonpert}

Here we demonstrate the effectiveness of the proposed renormalization prescription using the vertex functions of the vector operator, $\Gamma=\gamma^0$, which is free of mixing and is used for the calculation of physical matrix elements related to the unpolarized PDF. We checked that the conclusions drawn here are also valid for the axial and tensor operators. We use two ensembles of $N_f=2$ twisted mass fermions with a clover term, and Iwasaki gluons. The volume is $24^3\times48$, lattice spacing $a=0.0938$ fm, and the corresponding values of the pion mass are 235 MeV and 340 MeV. More details can be found in Refs.~\cite{Alexandrou:2015sea,Alexandrou:2019lfo}. We find that the level of improvement is the same for both ensembles, and present results for the 235 MeV case. It should be noted that the non-perturbative vertex functions have been obtained using five steps of stout smearing, while the perturbative calculation is performed without a smearing. Therefore, the effect of the subtraction of lattice artifacts can potentially be further improved when the stout dependence is calculated. We refer the reader to Sec.~\ref{sec:stout} for the calculational setup.

The effect of the improvement scheme is best evaluated after the $Z_{\cal O}$ estimates are converted to the $\overline{\rm MS}$ or ${\rm M}\overline{\rm MS}$ schemes, where the residual dependence on the initial RI scale is ${\cal O}(g^4)$.  Here, we use the ${\rm M}\overline{\rm MS}$ scheme~\cite{Alexandrou:2019lfo}, which is employed by the Extended Twisted Mass Collaboration (ETMC). All data have been evolved to a scale of 2 GeV using the conversion factor of Ref.~\cite{Constantinou:2017sej}.

As mentioned previously, the non-perturbative renormalization functions should ideally be determined on spatially isotropic momenta with $P4$ as close to 0.25, which correspond to the fully isotropic case (temporal and spatial). Such choices put very strict limits on the number of RI' scales in a range in which both hadronic contaminations are small and perturbation theory is valid. It is, thus, desirable to expand the range of momenta by relaxing the criterion for $P4$. Here, we use momenta that satisfy $0.25 \leq P4 \leq 0.41$, and have $(ap)^2$ up to 7.

\begin{figure}[ht!]
    \centering
    \includegraphics[scale=0.335]{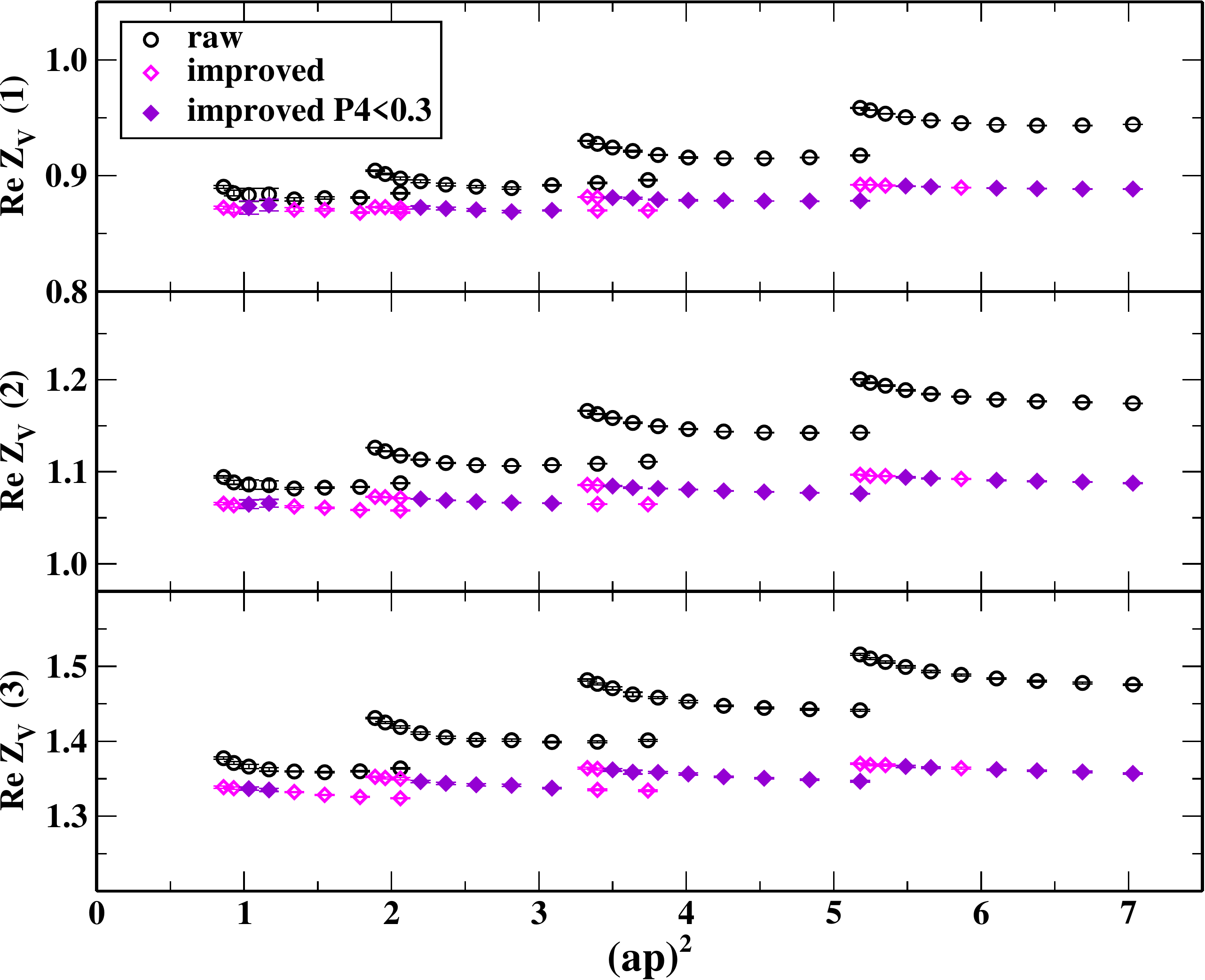}
    \includegraphics[scale=0.335]{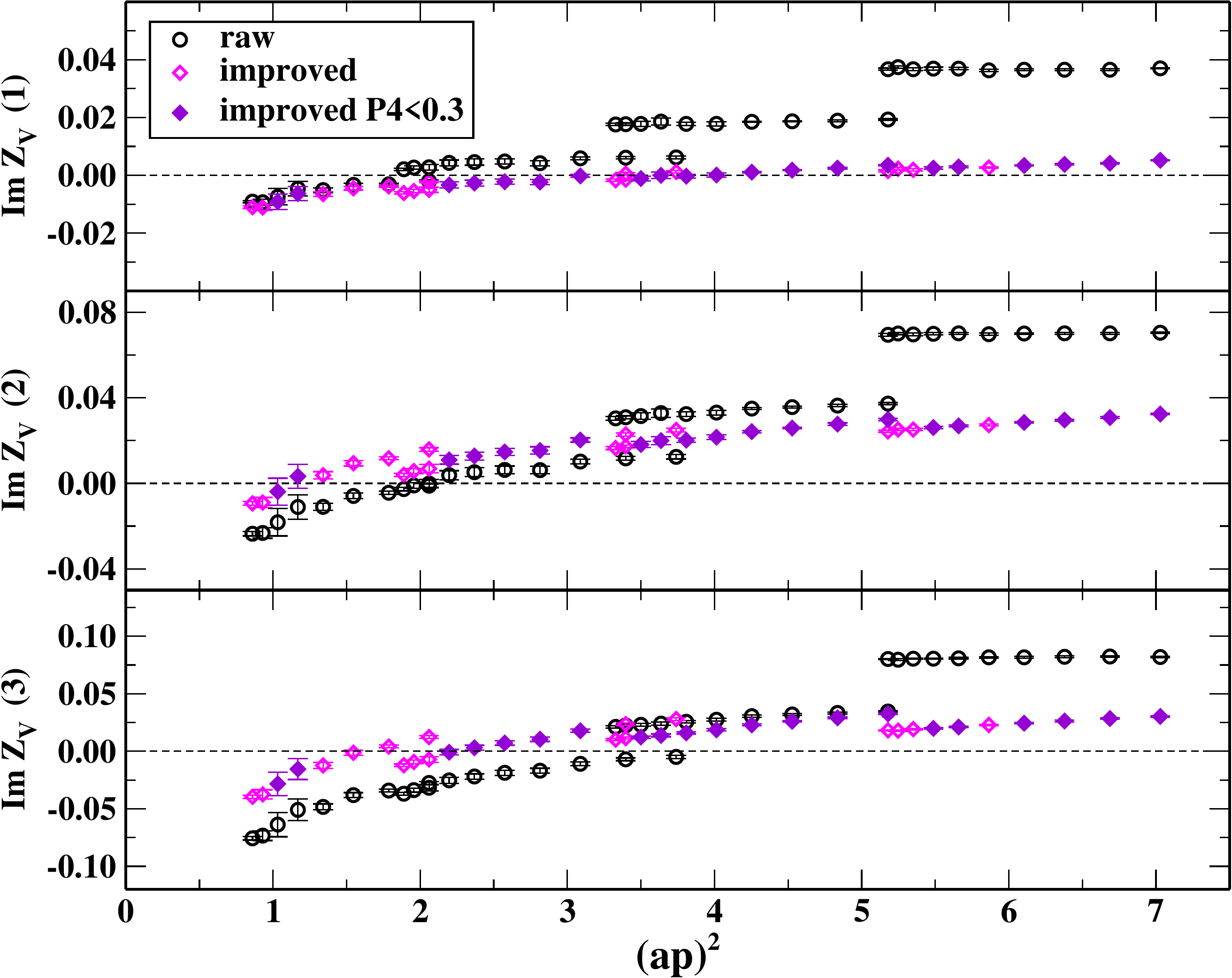}
    \vspace*{-0.3cm}
    \caption{Estimates of real (left) and imaginary (right) part of $Z_V(z/a)$ in the ${\rm M\overline{MS}}$ scheme at a renormalization scale of 2 GeV, as a function of the initial RI' scale. From top to bottom we show the values for $z=1,\,2,\,3$. Black circles correspond to Eq.~\eqref{renormZO} (raw), and magenta diamonds to Eq.~\eqref{renormZO_impr} (improved). Filled violet diamonds are the improved estimates of Eq.~\eqref{renormZO_impr} using RI' scales that satisfy $P4<0.3$\,.}
    \label{fig:Zv_123}
\end{figure}

\begin{figure}[ht!]
    \centering
    \includegraphics[scale=0.335]{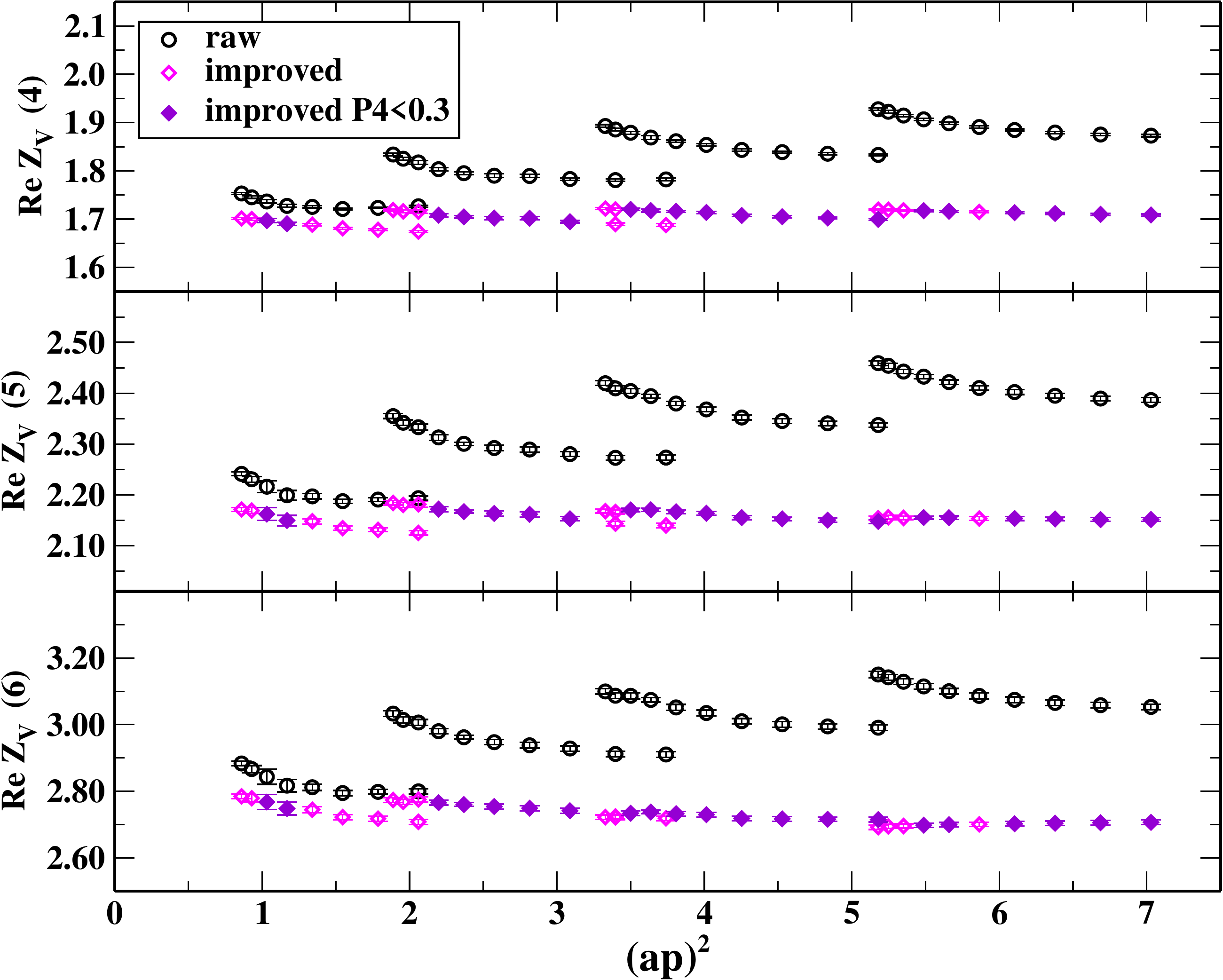}
    \includegraphics[scale=0.335]{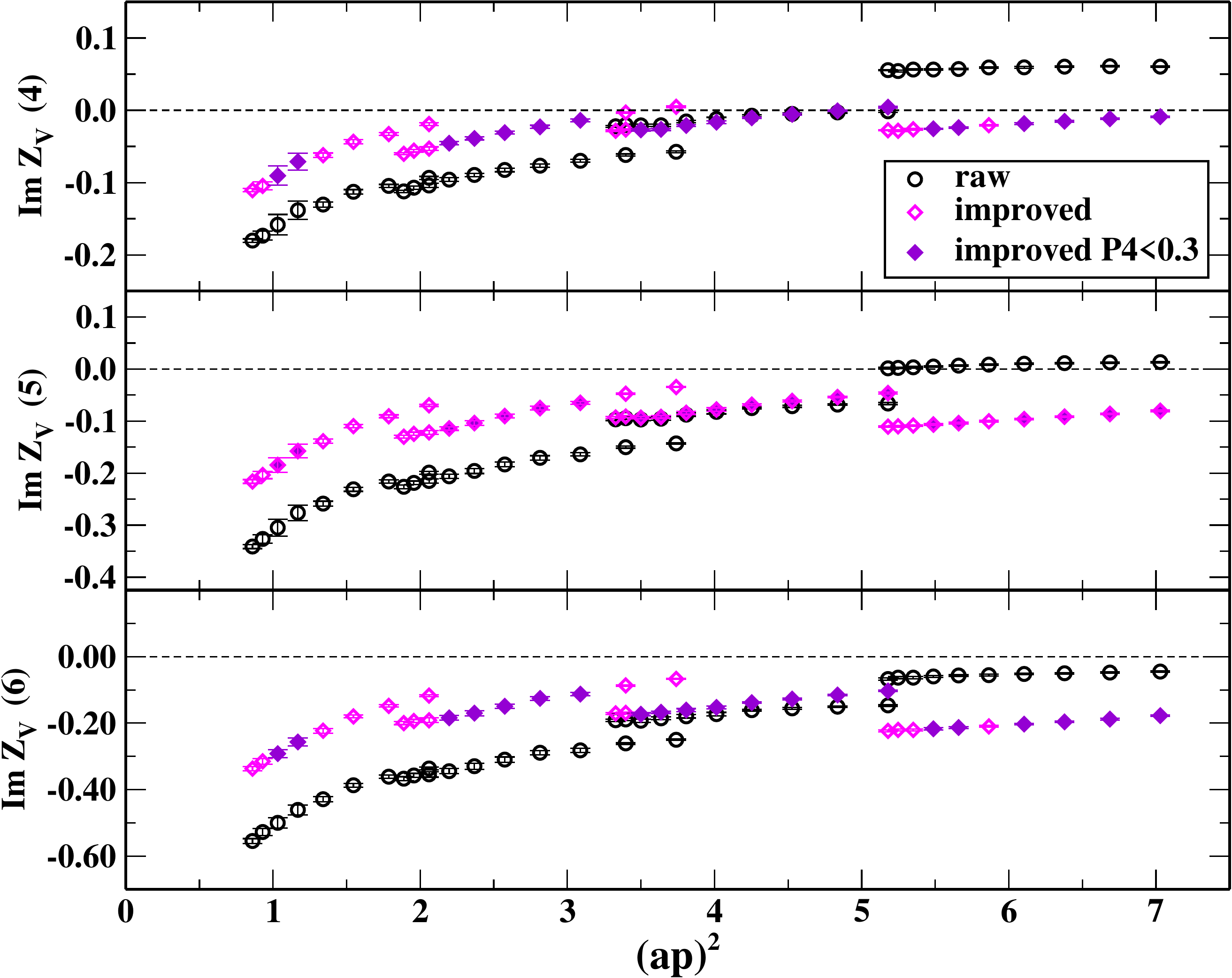}
        \vspace*{-0.4cm}
    \caption{Same as Fig.~\ref{fig:Zv_123} for $z=4,\,5,\,6$.}
    \label{fig:Zv_456}
\end{figure}

\begin{figure}[ht!]
    \centering
    \includegraphics[scale=0.335]{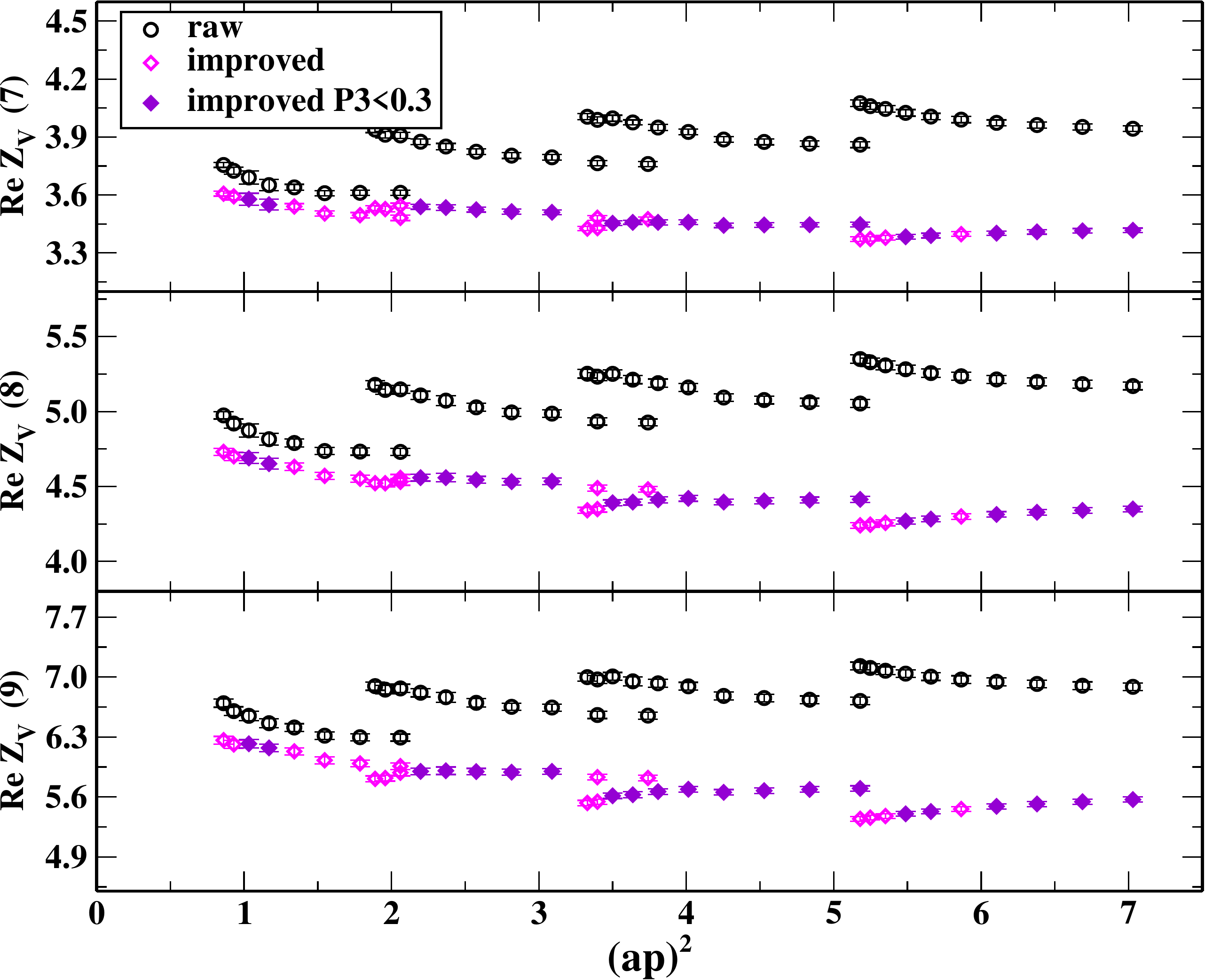}
    \includegraphics[scale=0.335]{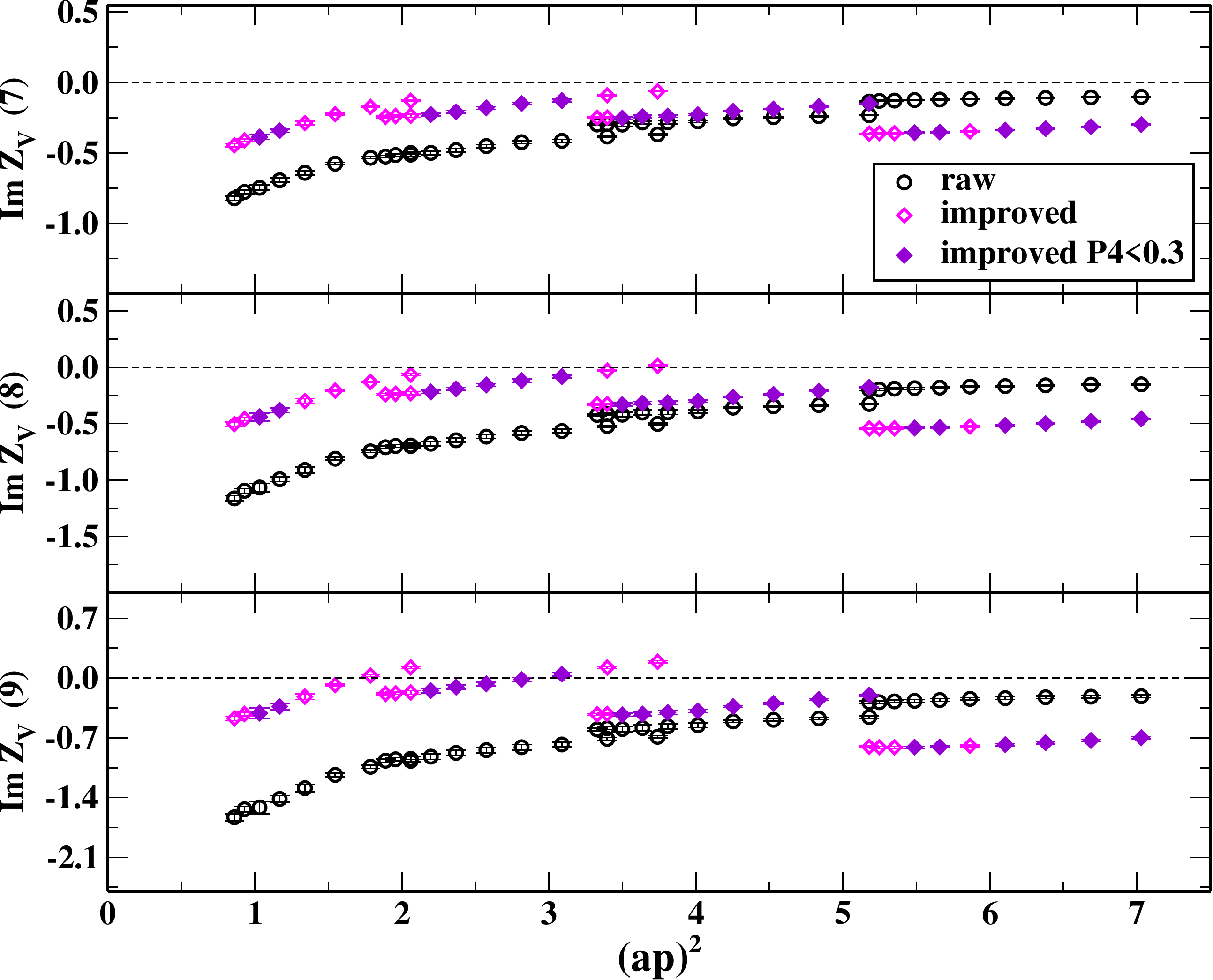}
        \vspace*{-0.3cm}
    \caption{Same as Fig.~\ref{fig:Zv_123} for $z=7,\,8,\,9$.}
    \label{fig:Zv_789}
\end{figure}

Fig.~\ref{fig:Zv_123} shows the raw non-perturbative data for $Z_V(1),\,Z_V(2),\,Z_V(3)$, as well as the corresponding improved ones. Similarly, Figs.~\ref{fig:Zv_456} - \ref{fig:Zv_789} show $Z_V$ for $z/a=4-6$ and $z/a=7-9$, respectively. Focusing on the real part of $Z_V$, we find that the effect of the subtraction of lattice artifacts as proposed in Eq.~\eqref{renormZO_impr} works very well up to $z = 7a \sim 0.65$ fm. Beyond that region, the subtraction method does not perform well, particularly for $P4>0.29$. This is not surprising because such values of z exceed the perturbative region, and the one-loop perturbative results become unreliable. Nevertheless, the method appears to be successful for $z<0.65$ fm. In particular, the ``wiggling'' effect between the different classes of momenta disappears for almost all values of $P4$. This is a highly nontrivial feature of the improved estimates. In addition, we observe the formation of good plateaux that allows one to reliably take the limit $(a\mu_0)^2\to 0$. The quality of the plateaux improves further if we apply a cut on the data at $P4=0.3$, as shown in the figures. It is noteworthy that the raw non-perturbative data produce a linear dependence in $(a\mu_0)^2$ for $P4=0.25$, which, however, has a sizeable slope leading to an unreliable estimate at $(a\mu_0)^2\to 0$. 

We also find interesting features in the imaginary part of $Z_V$. The first observation is that for $z \leq 4a$, the subtraction of lattice artifacts results in an estimate which is closer to zero than the purely non-perturbative ones. In fact, the imaginary part of the improved estimates is always closer to zero in the region $(a\mu_0)^2<5$. This is a desired feature because in dimensional regularization the poles are real to all orders in perturbation theory. The same feature should hold for non-perturbative calculations if a conversion factor to high enough order in $g^2$ were available. 
In addition, the improved estimates (Eq.~\eqref{renormZO_impr}) for the small-z region ($z<0.4$ fm) have a smoother dependence on $(a\mu_0)^2$ compared to the purely non-perturbative estimates (Eq.~\eqref{renormZO}). As $z$ increases, discontinuities are observed between different classes of momenta, in particular for $(a\mu_0)^2\gtrsim 3.5$. A milder effect is observed for momenta satisfying $P4 <0.3$. It is interesting to observe that the purely non-perturbative data have a smoother $(a\mu_0)^2$ dependence in the large-$z$ region compared to the improved ones.
This indicates that the real and imaginary parts have different contamination from finite-$a$ contributions. Thus, one can use an improvement scheme in which the subtraction of the artifacts can be applied differently in the real and imaginary parts. 
Also, including stout smearing in the perturbative calculation as described in Sec.~\ref{sec:stout}, has the potential to further improve the effect of the subtraction.

\vspace*{0.5cm}

\section{Stout Improvement}
\label{sec:stout}

A standard way of improving the behavior of operator matrix elements
is implemented by applying stout smearing to link
variables~\cite{Morningstar:2003gk}. This procedure is typically
performed as an iteration of smearing steps; starting from the
original link variables 
$U^{(0)}_\mu(x) \equiv U_{x,\, x+a\,\mu} \equiv {\rm exp}(i\,g\,a\,A^{(0)}_\mu(x+a\,\mu/2))$, the
$i^{\rm th}$ iteration step performs the following replacement~\cite{Horsley:2008ap}:
\be
U^{(i)}_\mu(x) \to U^{(i+1)}_\mu(x) = e^{i\,Q^{(i)}_\mu(x)}\, U^{(i)}_\mu(x)\,,
\ee 
where the definition of $Q_\mu(x)$ is [the superscript $(i)$, denoting
smearing step, is implicit]:
\be
  Q_\mu(x)=\frac{\omega}{2\,i} \left[V_\mu(x) U_\mu^\dagger(x) -
  U_\mu(x)V_\mu^\dagger(x) -\frac{1}{3} {\rm Tr} \,\Big(V_\mu(x)
  U_\mu^\dagger(x) -  U_\mu(x)V_\mu^\dagger(x)\Big)\right] \, .
\label{Q_def}
\ee
The ``stout coefficient'' $\omega$ is a tunable parameter
and $V_\mu(x)$ represents the sum over all staples associated with the
link $U_\mu(x)$.

The implementation of stout smearing on links which are
produced in numerical simulations is straightforward; several smearing
steps are generally applied in order to optimize improvement
in measured quantities. On the other hand,
calculations in lattice perturbation theory, in the presence of a
smeared action or of smeared operators, are prohibitively
complicated even at one loop, given the extraordinary proliferation of
terms in the corresponding vertices; as a result, such calculations
can usually be performed only when a maximum of two smearing steps
have been applied. \cite{Alexandrou:2016ekb,Constantinou:2013pba}.

Despite the sheer complexity of constructing and using vertices with
stout links, there are two general classes of vertices which can be
expressed in a compact form, even for an arbitrary number of smearing
steps: The first class regards vertices in which only one field is a
gluon. The second class regards vertices in which two fields are
gluons; this class leads to compact expressions only for that part of
the vertices which does not vanish when the color indices of the two
gluons are contracted (i.e. the part which is free of
commutators). Thus, there are several instances of Green's functions (GFs),
typically at one loop, which can be thoroughly computed with relative ease for
any number of smearing steps. Such instances include: (i) the fermion
propagator, (ii) GFs with external fermion-antifermion lines and an
insertion of any fermion local bilinear operator of arbitrary twist,
(iii) GFs with external fermion-antifermion lines and an insertion of
a fermion nonlocal bilinear. The latter instance is of course the
focus of the present work; however, the explanation which follows
regards generic one- and two-gluon vertices.

Let us first derive the terms up to ${\cal O}(g)$ (one-gluon) in a link after one smearing step. In terms of the Fourier transformed gluon field $\tilde{A}_\mu^{(0)}(p)$, one obtains:
\bea
U^{(1)}_\mu(x) &=& 1 + i\, g\, a\, \sum_{\rho=1}^4 \tilde{A}_\rho^{(0)}(p)\, e^{i\,p_\rho/2} \left( \delta_{\mu\rho} \left(1 + \omega \sum_{\sigma=1}^4 (e^{i\,p_\sigma} + e^{-i\,p_\sigma} -2)\right) + \omega (1 - e^{-i\,p_\rho} - e^{i\,p_\mu}+e^{i\,p_\mu-i\,p_\rho})\right) \nonumber \\
&=& 1 + i\, g\, a\, e^{i\,p_\mu/2} \,\sum_{\rho=1}^4 \tilde{A}_\rho^{(0)}(p)\, \left( \delta_{\rho\mu} - \omega (\delta_{\rho\mu} {\hat p}^2 - {\hat p}_\rho {\hat p}_\mu)\right)
\qquad\qquad \bigl[\,{\hat p}_\mu \equiv 2 \sin(p_\mu/2), \quad {\hat p}^2 \equiv \sum_{\rho=1}^4 {\hat p}_\rho^2\,\bigr] \nonumber \\
&\equiv& e^{\displaystyle i\, g\, a\, e^{i\,p_\mu/2} \tilde{A}_\mu^{(1)}(p)}, \quad {\rm where:} \quad \tilde{A}_\mu^{(1)}(p) = \sum_{\rho=1}^4 \tilde{A}_\rho^{(0)}(p)\, \left( (1-\omega\,{\hat p}^2) (\delta_{\rho\mu} - \frac{{\hat p}_\rho {\hat p}_\mu}{{\hat p}^2}) + \frac{{\hat p}_\rho {\hat p}_\mu}{{\hat p}^2}\right) + {\cal O}(g^2)
\eea
[For conciseness, an integral $\int_{-\pi}^\pi \, d^4p\, e^{i\,p\cdot {\bar x}}/(2\pi)^4$, accompanying each gluon field $\tilde{A}_\rho(p)$, is left implicit. ${\bar x}\equiv x/a$ is a four-vector with integer components.]

From the above expression it follows that the longitudinal part of the gluon field remains intact, while the transverse part is multiplied by a factor $(1-\omega\,{\hat p}^2)$. This feature propagates to successive smearing steps, with potentially different stout coefficients ($\omega$, $\omega'$):
\bea
\tilde{A}_\mu^{(2)}(p) &=& \sum_{\rho=1}^4 \tilde{A}_\rho^{(1)}(p)\, \left( (1-\omega'\,{\hat p}^2) (\delta_{\rho\mu} - \frac{{\hat p}_\rho {\hat p}_\mu}{{\hat p}^2}) + \frac{{\hat p}_\rho {\hat p}_\mu}{{\hat p}^2}\right) + {\cal O}(g^2)\nonumber\\
&=& \sum_{\rho=1}^4 \sum_{\sigma=1}^4 \tilde{A}_\sigma^{(0)}(p)\, \left( (1-\omega\,{\hat p}^2) (\delta_{\sigma\rho} - \frac{{\hat p}_\sigma {\hat p}_\rho}{{\hat p}^2}) + \frac{{\hat p}_\sigma {\hat p}_\rho}{{\hat p}^2}\right)\left( (1-\omega'\,{\hat p}^2) (\delta_{\rho\mu} - \frac{{\hat p}_\rho {\hat p}_\mu}{{\hat p}^2}) + \frac{{\hat p}_\rho {\hat p}_\mu}{{\hat p}^2}\right) + {\cal O}(g^2)\nonumber\\
&=& \sum_{\sigma=1}^4 \tilde{A}_\sigma^{(0)}(p)\, \left( (1-\omega\,{\hat p}^2) (1-\omega'\,{\hat p}^2) (\delta_{\sigma\mu} - \frac{{\hat p}_\sigma {\hat p}_\mu}{{\hat p}^2}) + \frac{{\hat p}_\sigma {\hat p}_\mu}{{\hat p}^2}\right) + {\cal O}(g^2)
\eea
Thus, a succession of $N$ steps with the same coefficient $\omega$ leads to:
\be
U^{(N)}_\mu(x) = e^{\displaystyle i\, g\, a\, e^{i\,p_\mu/2} \tilde{A}_\mu^{(N)}(p)}, \qquad
\tilde{A}_\mu^{(N)}(p) =  \sum_{\sigma=1}^4 \tilde{A}_\sigma^{(0)}(p)\, \left( (1-\omega\,{\hat p}^2)^N (\delta_{\sigma\mu} - \frac{{\hat p}_\sigma {\hat p}_\mu}{{\hat p}^2}) + \frac{{\hat p}_\sigma {\hat p}_\mu}{{\hat p}^2}\right) + {\cal O}(g^2)
\ee

Terms with two or more gluons in the exponent of smeared links (i.e., the ${\cal O}(g^2)$ terms in $\tilde{A}_\mu^{(N)}(p)$) are considerably more complicated. However, given that $U^{(N)}_\mu(x)$ is special unitary by construction, two-gluon terms in its exponent will necessarily involve a commutator; consequently, such terms will give a vanishing contribution in Feynman diagrams where the color indices of the two gluons end up being identified. In such diagrams, the two-gluon expression arising from a smeared link is simply:
\be
U^{(N)}_\mu(x) = \frac{(i\, g\, a)^2}{2}\, \bigl(e^{i\,p_\mu/2} \tilde{A}_\mu^{(N)}(p)\bigr)\, \bigl(e^{i\,p'_\mu/2} \tilde{A}_\mu^{(N)}(p')\bigr) \qquad {\rm [\,only\ 2{-}gluon\ part,\ no\  commutators\,]}
\ee

Let us apply the above to the zero-, one- and two-gluon vertices of the nonlocal operator $\mathcal{O}_\Gamma^{(N)}$, defined as in Eq.~\eqref{Oper}, with $N$-fold smeared links. Setting $z=a\,n$, $n\in \mathbb N$, we obtain:
\bea
\mathcal{O}_\Gamma^{(N)} &=& \int_{-\pi}^\pi \frac{d^4k\, d^4k'}{(2\pi)^8}\,
e^{i\,(k'{-}k)\cdot {\bar x}}\, e^{i\,k'_\mu\, n}\, {\tilde{\overline\psi}}(k)\, \Gamma\, V\, {\tilde\psi}(k'), \qquad {\rm where,\ omitting\ commutators:}\\
V &=& \Biggl[
{\mathbb{1}}+ \frac{g\, a}{2}\, \int_{-\pi}^\pi \frac{d^4p}{(2\pi)^4}\,
e^{i\,p\cdot {\bar x}}\sum_{\rho=1}^4 \tilde{A}_\rho^{(0)}(p)\, \frac{e^{i\,p_\mu\,n}-1}{\sin(p_\mu/2)} \, \left( (1-\omega\,{\hat p}^2)^N (\delta_{\rho\mu} - \frac{{\hat p}_\rho {\hat p}_\mu}{{\hat p}^2}) + \frac{{\hat p}_\rho {\hat p}_\mu}{{\hat p}^2}\right)\nonumber\\
&&\phantom{\Biggl[{\mathbb{1}}}+ \frac{(g\, a)^2}{8}\, \int_{-\pi}^\pi \frac{d^4p\,d^4p'}{(2\pi)^8}\,
e^{i\,(p+p')\cdot {\bar x}}\sum_{\rho,\sigma=1}^4  \tilde{A}_\rho^{(0)}(p)\,\tilde{A}_\sigma^{(0)}(p')\, \frac{e^{i\,p_\mu\,n}-1}{\sin(p_\mu/2)} \, \frac{e^{i\,p'_\mu\,n}-1}{\sin(p'_\mu/2)} \, \cdot \nonumber\\
&&\qquad\qquad\qquad\qquad \left( (1-\omega\,{\hat p}^2)^N (\delta_{\rho\mu} - \frac{{\hat p}_\rho {\hat p}_\mu}{{\hat p}^2}) + \frac{{\hat p}_\rho {\hat p}_\mu}{{\hat p}^2}\right)
\left( (1-\omega\,{\hat p}'^2)^N (\delta_{\sigma\mu} - \frac{{\hat p}'_\sigma {\hat p}'_\mu}{{\hat p}'^2}) + \frac{{\hat p}'_\sigma {\hat p}'_\mu}{{\hat p}'^2}\right)\Biggr]
\eea

The appearance of the stout parameter $\omega$ exclusively in the combination $(1-\omega\,{\hat p}^2)^N$ provides a point of reference for the numerical values of $\omega$ to be employed: Given that the maximum value of ${\hat p}^2$ is 16, and its average value is 8, one may expect that values of $\omega$ in the range $1/16 \lesssim \omega \lesssim 1/8$ tend to eliminate the transverse part of the gluon fields after several smearing steps. This, in particular, implies a vanishing result for diagrams d2, d3, d4 (see Fig.~\ref{fig1}) in the Landau gauge, leaving only diagram d1; since the latter coincides with its continuum counterpart, it follows that there would be no lattice corrections to this order. Conversely, values of $\omega$ beyond $\omega \gtrsim 1/8$ risk a disproportionate increase of lattice artifacts in the transverse gluon field. Non-perturbatively, the value of $\omega$ can be tuned using the criterion that the plaquette reaches maximum value for a given number of smearing steps, and typical values for $\omega$ are around 0.1.

As a demonstration, we show numerical results for the linear divergence
\begin{equation}
\Lambda_{d_4}(\omega) = \frac{g^2\, C_F}{16 \pi^2} \,\frac{|z|}{a}\,\left(e_0 + e_{N_\omega} \right)
\end{equation}
where the coefficient $e_{N_\omega}$ depends on the gluon action, the steps of stout smearing, $N_\omega$, and the value of $\omega$. We separate the coefficient of the linear divergence in the absence of stout smearing, $e_0$. Here we use Iwasaki gluons and find $e_0=-12.98$. The numerical values of $e_{N_\omega}$ are shown in Table~\ref{tab:statS} for a few choices of $\omega$ and $N_\omega \in [1,10]$. 
Given the gauge invariance of the linear divergence, the Landau-gauge statements of the previous paragraph, regarding dependence on $\omega$ and $N_\omega$ are directly applicable here.
\begin{table}[ht!]
\begin{center}
\begin{tabular}{|l|cccccccccc|c|}
\hline 
\,\,\backslashbox{$\omega$}{ $N_\omega$}  \,\,
&\hspace*{0.5cm} 1 \hspace*{0.5cm}
&\hspace*{0.5cm}2 \hspace*{0.5cm}
&\hspace*{0.5cm}3 \hspace*{0.5cm}
&\hspace*{0.5cm}4 \hspace*{0.5cm}
&\hspace*{0.5cm}5 \hspace*{0.5cm}
&\hspace*{0.5cm}6 \hspace*{0.5cm}
&\hspace*{0.5cm}7 \hspace*{0.5cm}
&\hspace*{0.5cm}8 \hspace*{0.5cm}
&\hspace*{0.5cm}9 \hspace*{0.5cm}
&\hspace*{0.5cm}10 \hspace*{0.5cm}\\
\hline 
\,\, $0.001$ \,\,
&\,\, 0.085\qquad 
&\,\,0.168\qquad 
&\,\,0.251\qquad 
&\,\,0.333\qquad 
&\,\,0.414\qquad 
&\,\,0.494\qquad 
&\,\,0.573\qquad 
&\,\,0.651\qquad 
&\,\,0.729\qquad 
&\,\,0.805\qquad \\
\hline
\,\, $0.010$ \,\,
&\,\,  0.825\qquad
&\,\,1.560\qquad 
&\,\,2.217\qquad 
&\,\,2.806\qquad 
&\,\, 3.335\qquad 
&\,\,3.812\qquad 
&\,\,4.243\qquad 
&\,\,4.633\qquad 
&\,\,4.988\qquad 
&\,\,5.311\qquad \\
\hline
\,\, $0.050$ \,\,
&\,\, 3.643 \qquad
&\,\,5.621\qquad 
&\,\,6.817\qquad 
&\,\,7.605\qquad 
&\,\,8.161\qquad 
&\,\,8.575\qquad 
&\,\,8.897\qquad 
&\,\,9.155\qquad 
&\,\,9.369\qquad 
&\,\,9.549\qquad \\
\hline
\,\, $0.100$ \,\,
&\,\, 6.083 \qquad
&\,\,7.849\qquad 
&\,\,8.716\qquad 
&\,\,9.246\qquad 
&\,\,9.613\qquad 
&\,\,9.887\qquad 
&\,\,10.10\qquad 
&\,\,10.28\qquad 
&\,\,10.42\qquad 
&\,\,10.55\qquad \\
\hline
\,\, $0.125$ \,\,
&\,\, 6.852 \qquad
&\,\,8.426\qquad 
&\,\,9.180\qquad 
&\,\,9.643\qquad 
&\,\,9.967\qquad 
&\,\,10.21\qquad 
&\,\,10.40\qquad 
&\,\,10.56\qquad 
&\,\,10.69\qquad 
&\,\,10.80\qquad \\
\hline
\,\, $0.150$ \,\,
&\,\,7.319\qquad
&\,\,8.805\qquad 
&\,\,9.500\qquad 
&\,\,9.927\qquad 
&\,\,10.22\qquad 
&\,\,10.45\qquad 
&\,\,10.62\qquad 
&\,\,10.77\qquad 
&\,\,10.89\qquad 
&\,\,10.99\qquad \\
\hline
\end{tabular}
\caption{Stout-step dependence of linear divergence, $e_{N_\omega}$, for Iwasaki gluon action and various values of $\omega$.}
\label{tab:statS}
\end{center}
\end{table}

\begin{figure}[ht!]
\centerline{\includegraphics[scale=0.34]{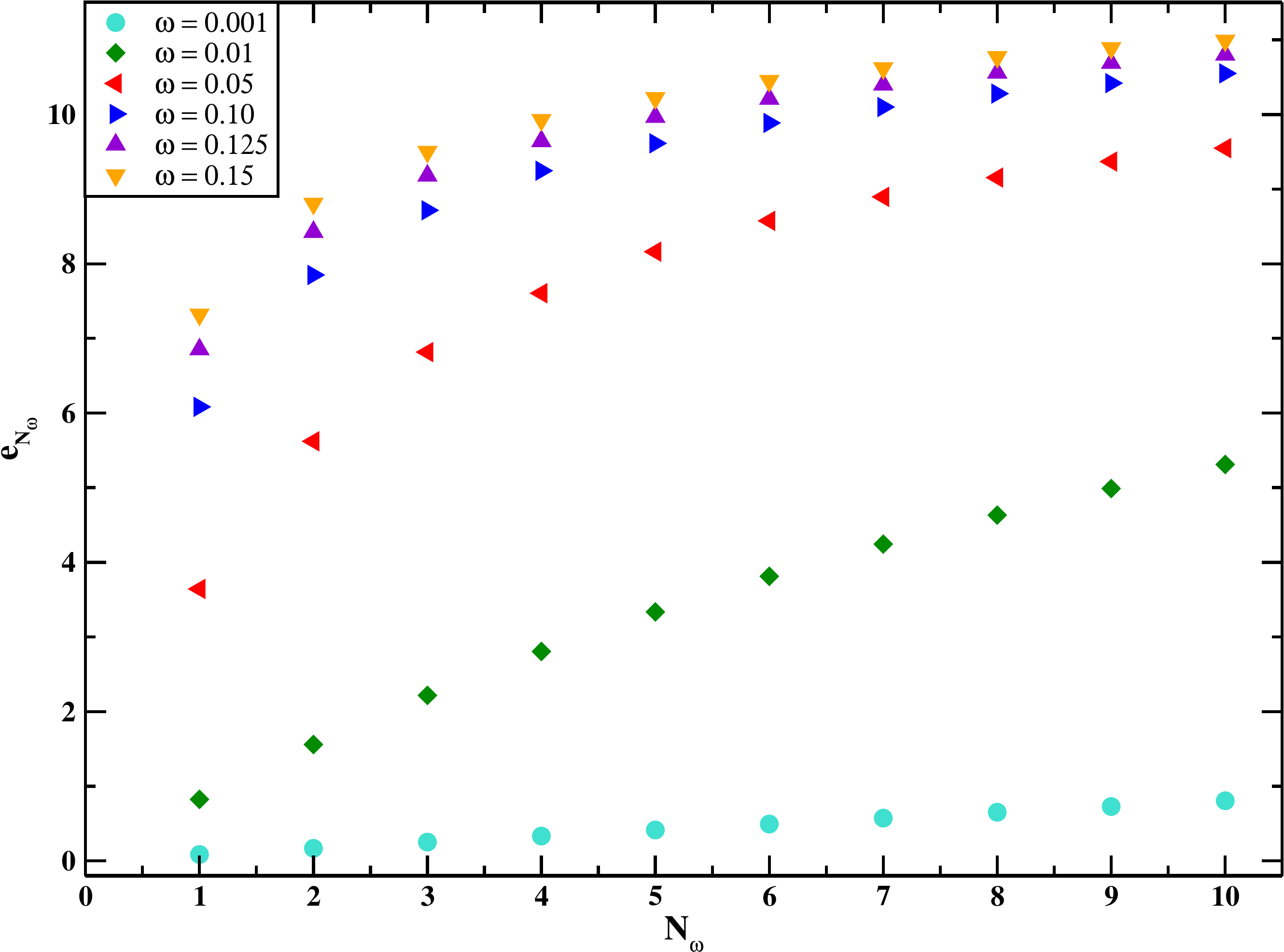}\,\,\,
\includegraphics[scale=0.34]{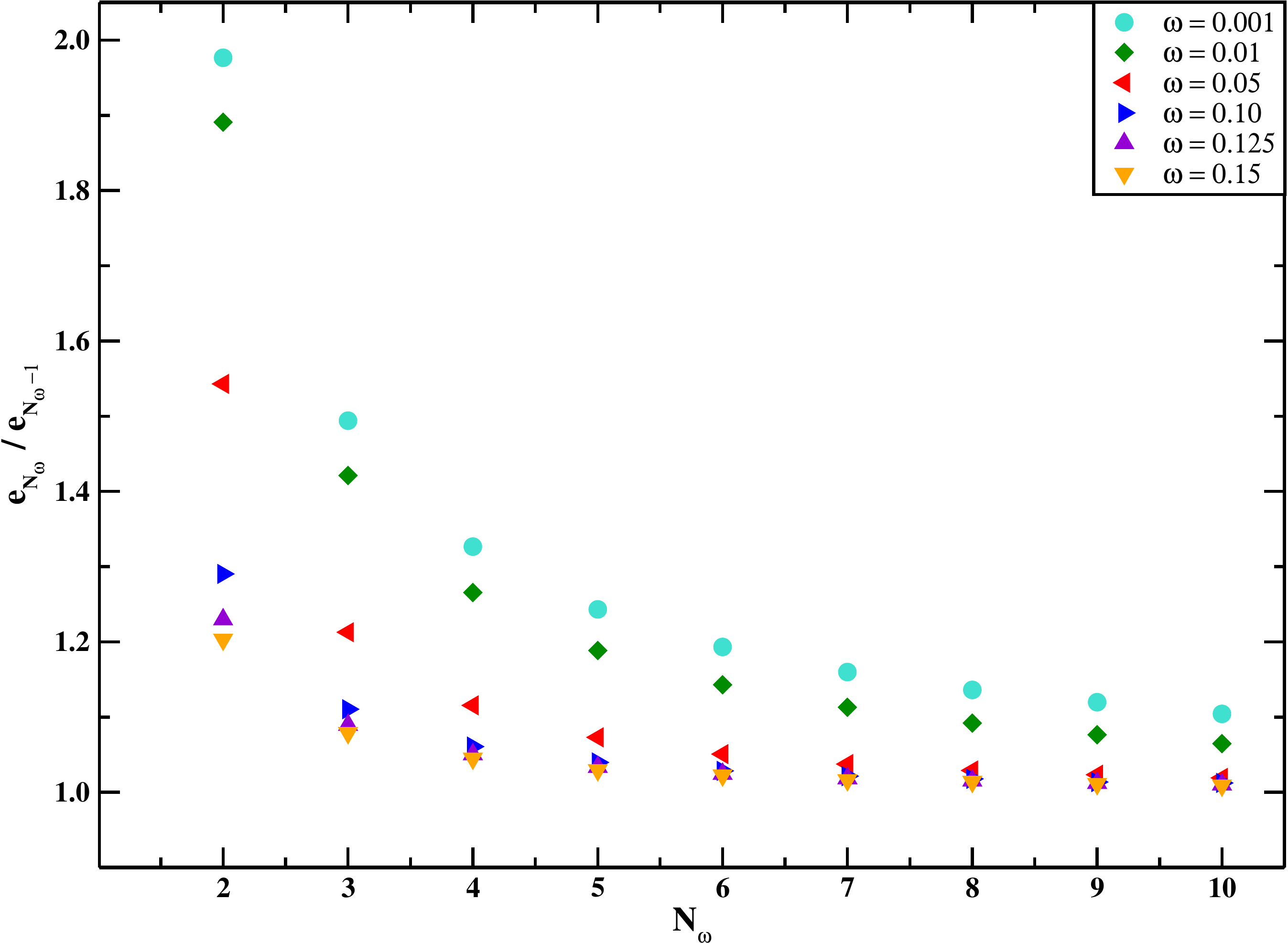}}
\caption{\small{Left: Stout-dependent coefficient, $e_{N_\omega}$, of the linear divergence in the Green's functions as a function of the stout steps, $N_\omega \in [1-10]$. The results for the stout smearing parameter, $\omega=0.001,\,0.01,\,0.05,\,0.1,\,0.125,\,0.15$ are shown with cyan circles, green diamonds, red left triangles, blue right triangles, purple up triangles, orange down triangles, respectively. Right: The ratio $e_{N_\omega}/e_{N_\omega-1}$ for $N_\omega \in [1-10]$. The notation is the same as in the left panel.}}
\label{fig:omega}
\end{figure}

\noindent
The values $\omega=0.1,\,0.125,\,0.15$ have been implemented in numerical calculation of matrix elements of nonlocal operators~\cite{Alexandrou:2019lfo,Chai:2020nxw,Alexandrou:2021oih}. The values $\omega < 0.1$ are included for pedagogical reasons. The data can be compared in different ways leading to various conclusions. Firstly, for all cases, the increase of stout steps leads to an increase of the $e_{N_\omega}$ value (see, also, left panel of Fig.~\ref{fig:omega}). Because of the different sign of $e_{N_\omega}$ and $e_0$, this leads to a suppression of the linear divergence contribution in the matrix elements.
This corroborates the use of multiple smearing steps in physical matrix elements to suppress the linear divergence in the renormalization functions. Secondly, very small values of $\omega$ lead to small additions to $e_0$, and, thus, the stout smearing does not have any practical benefit. Thirdly, $e_{N_\omega}$ is not too sensitive to the value of $\omega$ when chosen in the range $[0.1-0.15]$, which are the values employed in non-perturbative calculations of physical matrix elements. Lastly, the value of $e_{N_\omega}$ demonstrates convergence as the number of steps increases. This effect can also be seen in the right panel of Fig.~\ref{fig:omega} where we plot the ratio $N_\omega/N_{\omega-1}$. The ratio approaches unity, typically after five steps of smearing for $\omega =0.05 - 0.15$. While the setup and the conclusions described above are valid in one-loop perturbation theory, they are also indicative for the non-perturbative behaviour.

\section{Summary}

In this work, we propose an improvement to RI-type prescriptions for the renormalization of fermion bilinear nonlocal operators containing a straight Wilson line. The method is inspired by its successful implementation to local operators~\cite{Capitani:2000xi,Constantinou:2009tr,Gockeler:2010yr,Constantinou:2013ada, Alexandrou:2015sea}, and can be generalized to any operator, for instance, nonlocal operators with staple-shaped Wilson line, or gluon operators. The improved renormalization scheme is applied to non-perturbative vertex functions and uses results of the operators' Green's functions from a perturbative calculation in lattice QCD to all orders in the lattice spacing, ${\cal O}((g^2)^n\,a^\infty)$. The approach is applicable to any order in perturbation theory, but we find that the one-loop level ($n=1$) is sufficient to reduce the bulk of finite-$a$ effects. While the method utilizes perturbation theory, the renormalization functions maintain their non-perturbative nature: the subtraction of finite-$a$ effects as outlined in Eq.~\eqref{renormZO_impr} can be interpreted as an additional finite renormalization containing only higher orders in $a$. 

In our proof-of-concept calculation we implement the improved scheme to two $N_f=2$ ensembles of twisted mass fermions with a clover term, and Iwasaki gluons. The ensemble has $\beta=2.10$ ($a=0.0938$ fm) and lattice size $24^3\times48$. Here, we present results for the ensemble with a pion mass of 235 MeV, which was used in Ref.~\cite{Alexandrou:2019lfo} for the non-perturbative renormalization of PDFs. We note that our existing perturbative calculation is easily adaptable to any gluon action (Plaquette, tree-level Symanzik, Iwasaki and tadpole-improved L\"uscher-Weisz (TILW)), and Wilson/clover fermions. Also, the Dirac structure of the operator, $\Gamma$, and the parameters $g^2$, $c_{\rm SW}$, the gauge fixing parameter $\alpha$, $(ap)^2$ are chosen at the last stage of the calculation. 

We find that the proposed renormalization scheme (Eq.~\eqref{renormZO_impr}) is advantageous compared to the purely non-perturbative one (Eq.~\eqref{renormZO}), as it has better control of unwanted finite-$a$ effects. As a consequence, one can include in non-perturbative calculations a wider range of renormalization scales, $(a\mu_0)^2$, as well as $P4$ values beyond the most democratic ones ($P4=0.25$), a measure of the Lorentz non-invariant finite-$a$ contributions. For example, in the calculations of the renormalization of nonlocal operators done so far~\cite{Alexandrou:2017huk,Alexandrou:2018pbm,Alexandrou:2018eet,Alexandrou:2019lfo,Chai:2020nxw,Bhattacharya:2020xlt,Alexandrou:2020zbe,Alexandrou:2020uyt,Bhattacharya:2021moj,Alexandrou:2021bbo} only about 10 values of the scale were used covering a small $(a\mu_0)^2$ interval , typically $[2-4]$; These momenta are chosen based on the $P4<0.28$ criterion. Here, we are able to include in our non-perturbative analysis 38 values of  $(a\mu_0)^2$ that cover a range of about $[1-7]$. The improvement of the renormalization functions works well even at $P4$ around 0.4 for $z\le6$, which is a significant advantage. However, caution is needed because of the limitations of perturbation theory that are manifested at high values of $z$. In this analysis we find improvement for $z$ up to 0.65 fm. Another conclusion from this calculation is that the subtraction of ${\cal O}(g^2 a^\infty)$ terms has different effectiveness in the real and imaginary parts of the renormalization function with respect to $z$. In particular, the real part is still improved for $z>0.65$ fm, while the imaginary part worsens. With this in mind, one can employ an alternative scheme to subtract the artifacts, which has different $z_{\rm max}$ for the real and imaginary parts.

As mentioned in the main text, the subtraction of the finite-$a$ terms is done in the absence of stout smearing, which is computationally more expensive due to a fast increase of additional terms in the Green's functions. In this work we lay the foundation for a calculation that contains stout smearing. Our proposal gives, for the first time, a prescription to include an arbitrary number of stout steps with the same computational cost as a single step. This is applicable to cases involving Feynman diagrams with up to two gluons that will be contracted to each other (see, e.g., d4 in Fig.~\ref{fig1}), like the cases presented here. This is very powerful, as the non-perturbative calculations are performed with 5-20 stout steps, which is prohibitive in lattice perturbation theory without the approach of Sec.~\ref{sec:stout}.

The applicability of the method can encompass a wide range of operators for both quark and gluon external fields, local and nonlocal cases. For gauge-invariant nonlocal operators, other choices of Wilson lines can be accommodated. Perturbative and non-perturbative results exist in the literature, such as nonlocal operators with staple-shaped Wilson lines~\cite{Constantinou:2019vyb,Shanahan:2019zcq,Zhang:2022xuw} and could be an interesting extension of this work. Another direction that can be immediately pursued is the incorporation of the improved RI-scheme within the hybrid scheme~\cite{Ji:2020brr}, in which an RI-type prescription can be employed in the small-$z$ region.

\begin{acknowledgements}
M.C. acknowledges financial support by the U.S. Department of Energy Early Career Award under Grant No.\ DE-SC0020405. 
H.P. acknowledges financial support from the projects ``Quantum Fields on the Lattice'' and ``Lattice Studies of Strongly Coupled Gauge Theories: Renormalization and Phase Transition'', funded by the Cyprus Research and Innovation Foundation under contract numbers EXCELLENCE/0918/0066 and EXCELLENCE/0421/0025, respectively. The computations of the non-perturbative vertex functions were carried out on facilities of the USQCD Collaboration, which are funded by the Office of Science of the U.S. Department of Energy.

\end{acknowledgements}

\bibliography{references.bib}

\end{document}